\documentclass[aps,prd,twocolumn,superscriptaddress,groupedaddress,preprintnumbers]{revtex4} 
\usepackage{graphicx}  
\usepackage{dcolumn}   
\usepackage{bm}        
\usepackage{amssymb}   
\usepackage{amsmath}
\usepackage{mathtools}
\usepackage{cases}
\usepackage{mathrsfs}
\usepackage{color}
\usepackage{enumitem}
\usepackage[normalem]{ulem}
\newcommand{\mpl}{M_{\text{Pl}}}

\newcommand{\be}{\begin{equation}}
\newcommand{\ee}{\end{equation}}
\newcommand{\bea}{\begin{eqnarray}}
\newcommand{\eea}{\end{eqnarray}}
\newcommand{\nn}{\nonumber}

\newcommand{\rh}{\text{rh}}

\begin{document}
	\preprint{IPMU19-0138}
	\preprint{YITP-19-91}

	\title{Universal Upper Bound on the Inflationary Energy Scale from\\
	\vspace{0.3em}
	 the Trans-Planckian Censorship Conjecture}
\author{ { Shuntaro Mizuno$^{1}$}, {Shinji Mukohyama$^{2,3}$},  {Shi Pi$^{3,4}$}, {Yun-Long Zhang$^{2}$}\\
\it  {${}^1$
Department of Liberal Arts and Engineering Sciences,\\
National Institute of Technology, Hachinohe College, Aomori, 039-1192, Japan}
\it  {$^{2}$Center for Gravitational Physics, Yukawa Institute for Theoretical Physics(YITP)}, \\
\it  {Kyoto University, Kyoto 606-8502, Japan} \\
\it   $^{3}$Kavli Institute for the Physics and Mathematics of the Universe (WPI), \\
\it  {The University of Tokyo Institutes for Advanced Study,}\\
\it  {The University of Tokyo, Kashiwa, Chiba 277-8583, Japan}\\
\it  $^{4}${CAS Key Laboratory of Theoretical Physics, Institute of Theoretical Physics,}\\
\it  {Chinese Academy of Sciences, Beijing 100190, China}
}
	\date{\today}
	\begin{abstract}
	We study the constraint on the inflationary energy scale from the recently proposed Trans-Planckian Censorship Conjecture (TCC). We find a universal upper bound on the inflationary Hubble expansion rate $H_\text{inf}$ which is solely determined by the reheating temperature $T_\rh$: $H_\text{inf}/\mpl\lesssim T_0/T_\rh$, where $T_0$ is the photon temperature today. The upper limit can be saturated by a post-inflationary oscillatory stage with the critical equation-of-state parameter $w\approx-1/3$, or by an inflation model with multiple stages. In the lowest reheating temperature required for big bang nucleosynthesis, the upper bound on the tensor-to-scalar ratio $r$ at the CMB scales is $r\lesssim10^{-8}$, which can be realized in many string-inspired inflation models.
	\end{abstract}
	\maketitle

	\noindent	
\section{Introduction}\label{sec:intro}

One of the main goals of cosmology is to study the origin of the initial condition for the cosmological perturbations which source the cosmic microwave background radiation (CMB) anisotropies and the large scale structures we observe today. By WMAP and Planck observations~\cite{Hinshaw:2012aka,Aghanim:2018eyx}, we already know that the primordial curvature perturbation on scales larger than 1Mpc should be highly Gaussian and of order $10^{-5}$. These perturbations can originate from the quantum fluctuations during an accelerated expansion in the very early universe, i.e. inflation~\cite{Brout:1977ix,Sato:1980yn,Guth:1980zm,Starobinsky:1980te,Linde:1981mu,Albrecht:1982wi,Starobinsky:1979ty,Mukhanov:1981xt}. 
For reviews of the cosmological perturbation theory, see e.g. Refs.\cite{Kodama:1985bj,Mukhanov:1990me}.

It is believed that general relativity is a low energy effective field theory of the yet unknown quantum theory of gravity, of which the energy scale is the Planck scale $\mpl\equiv(8\pi G)^{-1/2}\approx2.435\times10^{18}~\text{GeV}$. Although we are still far away from establishing a successful quantum theory of gravity, some aspects that this unknown theory is expected to possess have been proposed. For instance, it is conjectured in \cite{Brennan:2017rbf,Obied:2018sgi} that effective field theory with a de Sitter vacuum is not compatible with the quantum theory of gravity, therefore it is in the ``swampland''. For a review of this swampland conjecture, see \cite{Palti:2019pca}. 

Following this logic, recently Ref.~\cite{Bedroya:2019snp}  
proposed the \textit{Trans-Planckian Censorship Conjecture} (TCC):
\textit{Any inflationary model which can stretch the quantum fluctuations with wavelengths smaller than the Planck scale out of the Hubble horizon is in the swampland}. 
Written in terms of the $e$-folding number $N$ by which the universe has expanded during inflation, it is 
\be\label{tcc}
e^N<\frac{\mpl}{H_e},
\ee
where $H_e$ is the Hubble expansion rate at the end of inflation. A companion paper, Ref.~\cite{Bedroya:2019tba}, has applied this TCC to inflation, and concluded that the Hubble expansion rate during inflation $H_\text{inf}$ must be as small as $10^{-1}\text{GeV}$, which predicts a completely negligible amplitude of primordial gravitational waves and implies that the initial condition of slow-roll inflation must be extremely fine-tuned. For discussions based on this result, see Refs.\cite{Cai:2019hge,Tenkanen:2019wsd,Das:2019hto}.

However, in the argument of \cite{Bedroya:2019tba}, two key assumptions are made: (1) the Hubble expansion rate does not change much during inflation and (2) the universe reheats instantaneously right after inflation. If we relax one or both of these conditions, it is possible to have a much larger upper bound.  For instance, the TCC 
constrains $H_e$, but the Hubble expansion rate in the early part of inflation which determines the energy scale of inflation and the tensor-to-scalar ratio $r$ that we observe/constrain today by experiments is much less constrained. This at the same time requires a low reheating temperature, which can be as low as $1\text{MeV}$ required by big bang nucleosynthesis. 
Motivated by this consideration, we study the TCC in inflation models with low reheating temperature and find that a more general and universal upper bound on the inflationary energy scale which is inversely proportional to the reheating temperature. 


 \section{Diagrammatic study of TCC}\label{sec:pic}
In Ref.~\cite{Bedroya:2019tba}, by assuming that the reheating happens right after the inflation, Bedroya, Brandenberger, Loverde and Vafa (BBLV) have derived from the TCC an upper bound on the inflationary energy scale which, when written in the Hubble expansion rate, gives $H_\text{inf}\lesssim10^{-20}\mpl$. Therefore an extremely small primordial GWs with tensor-to-scalar ratio $r\lesssim10^{-30}$, is a direct consequence of the TCC. However, as we commented above, when relaxing their two assumptions, we can significantly loosen this stringent bound, and reach a more general and universal bound which is inversely proportional to the reheating temperature. In this section, we estimate this new bound by a spacetime diagram of the expansion history.

\begin{figure}[htbp]
\begin{flushleft}
\includegraphics[trim={3cm 0.5cm 1cm 1cm}, clip, width=0.5\textwidth]{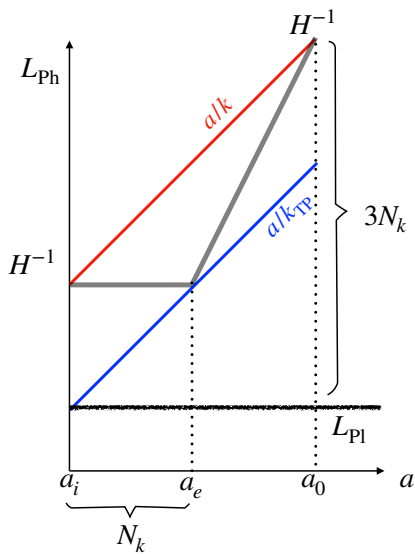}
\caption{Schematic spacetime diagram of the expansion of our universe from the beginning of inflation to today. For simplicity we have omitted the late matter- and dark-energy-dominated era. When the universe reheats instantaneously right after inflation as is studied in Ref.~\cite{Bedroya:2019tba}, it is very easy to derive $N\approx46$ and $H\approx10^{-20}\mpl$.} 
\label{fig:BBLV}
\end{flushleft}
\end{figure}

In Fig.~\ref{fig:BBLV}, with $N_k$ to be the $e$-folding 
number during inflation, we draw the history of the Hubble horizon and some characteristic wavelengths of our universe, and the trans-Planckian mode corresponding to the BBLV bound in \cite{Bedroya:2019tba} is also displayed. As we shall discuss later, the analysis of BBLV ignores contributions from the pre-inflationary epoch. For simplicity, we also neglect the late evolution of matter- and dark-energy-dominated era, which will be included in the analysis in the next section. The TCC forbids the trans-Planckian mode (drawn in blue line) to exceed the Hubble horizon (drawn in thick gray line)  during inflation, which gives the following relation for the critical case from the trigonometry:
\be
3N_k=\ln\frac{\mpl}{H_0}\approx138.7,
\ee 
which gives $N_k=46.2$. Again by using the TCC we have
\begin{align}
&\frac{H_\text{inf}}{\mpl}<e^{-N_k}=8.4\times10^{-21},\\\label{BBLV}
&r=\frac{2}{\pi^2\mathcal{P_R}}\left(\frac{H_\text{inf}}{\mpl}\right)^2<6.8\times10^{-33},
\end{align}
where we have used $\mathcal{P_R} \approx 2.1 \times 10^{-9}$ from Planck 2018~\cite{Aghanim:2018eyx}.
Eq.~(\ref{BBLV}) is just the BBLV bound given in \cite{Bedroya:2019tba}.

\begin{figure}[htbp]
\begin{flushleft}
\includegraphics[trim={2cm 0.5cm 1cm 1cm}, clip, width=0.5\textwidth]{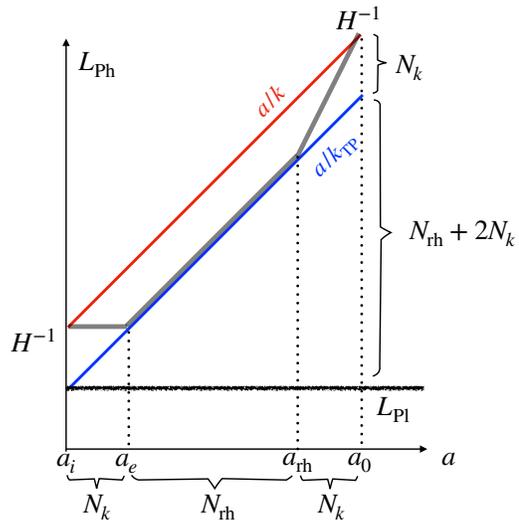}
\caption{There is a near-critical expansion after inflation with $w\approx-1/3$. In this case the upper limit on $H_k$ is determined by a simple relation $H_k/\mpl=T_0/T_\mathrm{rh}$.}
\label{fig:MMPZ}
\end{flushleft}
\end{figure}

If there is a stage when inflaton oscillates around its origin after inflation, which lasts for some $e$-folds $N_\rh$, we may expect this picture to change. In Fig.~\ref{fig:MMPZ}, we depict a near-critical case when the effective equation of state during the oscillatory stage is $w\approx-1/3$. The scale tangent to the Hubble horizon which expands from the Planck scale is drawn as the blue line, which by requiring the TCC gives the following relation for the critical case
\be
N_\rh+3N_k=\ln\frac{\mpl}{H_0}\approx138.7.
\ee
The inflationary $e$-foldings is equal to $\ln(a_0/a_\rh)$, which gives the following estimation
\be
N_k=\ln\frac{a_0}{a_\rh}\approx\ln\frac{T_\rh}{T_0},
\ee
where we again neglect all the numbers like the relativistic degrees of freedom, etc. When combined with the TCC, this gives us a bound for the inflationary Hubble horizon,
\be\label{bound1}
\frac{H_k}{\mpl}\lesssim\frac{T_0}{T_\text{rh}}.
\ee
This bound is the most general and universal bound we can derive from the TCC, which is the main result of our paper. The low reheating temperature considered here is crucial comparing to the BBLV bound in \cite{Bedroya:2019tba}, which is reproduced from \eqref{bound1} by taking $T_\rh\sim\sqrt{\mpl H_k}$. Eq.~\eqref{bound1} can be easily translated to the bound on $V$ or $r$, as we shall show in the following section.

\section{$e$-foldling counting and the upper bound for $r$}\label{sec:efold}

The universal bound \eqref{bound1} we derived is only an estimation, as we neglect the change of the relativistic degrees of freedom, and most importantly, the matter-dominated epoch of the universe. Roughly speaking, \eqref{bound1} can only guarantee that the modes of Planck length from the moment when $k_\text{eq}$-mode exit the horizon is kept subhorizon. There are still around 7 $e$-folds to be added from the largest comoving scale to $k_\text{eq}^{-1}$. To get the accurate bound we should count the $e$-folding number from the horizon exit of the largest comoving length $k_\text{DE}^{-1}$, which should be taken as the wavelength that reenters the horizon at the matter-dark-energy equality at $z\approx0.78$.  In this section, we will do $e$-folding counting following~\cite{Liddle:2003as,Martin:2010kz,Dai:2014jja,Munoz:2014eqa,Gong:2015qha,Cai:2015soa,Cook:2015vqa}, and obtain a more accurate bound.

Inflation assures that the largest scales we observe today were well inside the Hubble horizon during inflation. We consider a comoving wavenumber $k$ which exits the Hubble horizon during inflation at $k = a_kH_k$. The ratio of this scale to the scale of the current horizon, $a_0H_0$, can be written as
\begin{align}\nn
\ln \frac{k}{a_0H_0}&= \ln \left(\frac{a_k}{a_e} \frac{a_e}{a_\rh} \frac{a_\rh}{a_0} \frac{H_k}{H_0}\right) \\\label{history}
&= -N_k - N_\rh + \ln \frac{a_\rh}{a_0} + \ln \frac{H_k}{H_0} ,
\end{align}
where we have inserted some intermediate scale factors at different times in the cosmic evolution, like the end of inflation $a_e$ and the moment of instantaneous reheating $a_\rh$. During inflation, the universe has expanded for $N_k$ $e$-folds from the horizon exit of $k$-mode to the end of inflation. We assume, for simplicity, that the Hubble expansion rate is nearly a constant during inflation, which is favored by the recent observations of Planck that the inflationary potential is concave at 99\% confidence level~\cite{Aghanim:2018eyx}. The large-field region of a potential is a flat plateau, but near the origin it is power-law $\sim \phi^p$ with $p>1$ to cease inflation. 
This kind of potential can be realized in, for instance, the Starobinsky model~\cite{Starobinsky:1980te} or the $\alpha$-attractor~\cite{Kallosh:2013xya,Kallosh:2013hoa,Kallosh:2013yoa,Cai:2014bda,Maeda:2018sje}, which have attracted much attention as their predictions of tensor-to-scalar ratio $r$ and the spectral tilt $n_s$ are close to the sweet spot of the Planck contour. 
When oscillating around the origin of the potential, the inflaton still dominates the energy density, which gives an effective equation-of-state parameter $w$ after averaging over many periods $w=(p-2)/(p+2)$~\cite{Turner:1983he}. Then the energy density decays as $\rho\propto a^{-3(1+w)}$, until the time when the Hubble expansion rate drops down to become comparable to the decay rate of the inflaton to the standard model or intermediate stage particles. At that moment, reheating happens and the energy density of the universe transfers from the inflaton to the thermalized particles. We suppose that the reheating process is instantaneous, after which the universe is in thermal equilibrium with a temperature of $T_\rh$. After reheating the universe evolves as the usual hot big bang universe with $\rho_r\propto a^{-4}$. Then, the $e$-folding number for which the oscillation lasts can be written as
\begin{equation}\label{Nrh}
N_\rh\equiv\ln\frac{a_\rh}{a_e}=\frac{1}{3(1+w)} \ln\left( \frac{\rho_e}{\rho_\rh} \right) \, ,
\end{equation}
where $\rho_e=3\mpl^2H_e^2$ is the energy density at the end of inflation, and $\rho_\rh=g_{*}(T_\rh)\pi^2T_\rh^4/30$ is the energy density when the universe becomes thermalized, with $g_*(T_\rh)$ the effective degrees of freedom for the energy density at $T_\rh$. For standard model,  $g_*(T_\rh)$ varies from 10.75 at $T_\rh\approx1~\text{MeV}$ to  $106.75$ at $T_\rh\approx170~\text{GeV}$~\cite{Husdal:2016haj}. In $\alpha$-attractor-like models, the energy density at the end of inflation does not change much during inflation, which can be roughly estimated by the energy density at the horizon exit of $k$-mode, i.e. $\rho_e\approx\rho_k=3\mpl^2H_k^2$. 
The expansion after the universe being thermalized, $a_\rh/a_0$ can be easily derived by the entropy conservation in the thermal universe~\cite{Kolb:1990vq},
\begin{equation}\label{Trh}
\frac{a_\rh}{a_0} = \left(\frac{g_{*s}(T_\rh)}{g_{*s}(T_0)} \right)^{-1/3} \frac{T_0}{T_\rh} \, .
\end{equation}
If there is huge entropy generation in the evolution of the universe, \eqref{Trh} does not hold, yet our main result will not be altered. We then substitute \eqref{Trh} together with \eqref{Nrh} to \eqref{history}, to have
\begin{align}\nn
N_k&=\ln\frac{T_0}{H_0}-\frac{2}{3(1+w)}\ln\frac{H_e}{H_k}\\\nn
&+\frac{3w-1}{3(w+1)}\ln\frac{\mpl}{T_\rh}+\frac{1+3w}{3(w+1)}\ln\frac{H_k}{\mpl}\\\label{Nk}
&-\ln\frac k{a_0H_0}+\frac13\ln\left[\left(\frac{g_*(T_\rh)\pi^2}{90}\right)^{\frac1{1+w}}\frac{g_{*s}(T_0)}{g_{*s}(T_\rh)}\right].
\end{align}
We take the best-fit of Planck 2018 values~\cite{Aghanim:2018eyx}, with the relativistic degrees of freedom 
$g_{*}(1\text{MeV})=g_{*s}(1\text{MeV})=10.75$ and $g_{*s}(T_0)=43/11$~\cite{Husdal:2016haj}. As for our future use, we calculate the third line of \eqref{Nk} by taking its lower limit at $g_*=g_{*s}=10.75$ and $w=-1/3$.  The difference of taking different values is always of order $\mathcal O(0.1)$, so can be neglected. As we discussed, $k$ must be the largest comoving scale that ever enters the horizon, which is the cosmological horizon at the dark-energy-matter equality at $a_\text{DE}/a_0\approx0.78$, which gives $\ln(k_\text{DE}/a_0H_0)=\ln(a_\text{DE}/a_0)\approx-0.25$. Then we have accidental cancellation that the last line of \eqref{Nk} is $\mathcal{O}(10^{-3})$ 
and can be safely neglected, which gives
\be\label{Nk2}
N_k=\ln\frac{T_0}{H_0}+\frac{3w-1}{3(w+1)}\ln\frac{\mpl}{T_\rh}+\frac{1+3w}{3(w+1)}\ln\frac{H_k}{\mpl}.
\ee
We already take $H_k=H_e$ as we assume there is no magnificent change of $H$ during inflation. 
Because of the oscillation stage and the low reheating temperature, the $e$-folding number during inflation $N_k$ can be much shorter than $\mathcal{O}(60)$ if $w<1/3$. To compare this with the TCC, we first note that the sub-Planckian modes are only possible to be stretched out of the horizon during an accelerated expansion, i.e. inflation. Therefore in the TCC inequality \eqref{tcc}, $N$ is the $e$-folding numbers of inflation, which is our $N_k$ if we pick $a_0/k$ as the largest scale ever reenter the horizon in the current epoch. Then the TCC condition is
\be\label{tcc2}
N_k<\ln\frac{\mpl}{H_e}.
\ee
It involves the Hubble expansion rate at the end of inflation which, by our assumption, is the same as $H_k$. The subtlety and extension of choosing a different $H_e$ will be discussed in the next section, where a group of examples with more complicated intermediate stages will be found to predict the same upper bound. Here we have
\be\label{Nk3}
 \ln\frac{H_k}{\mpl} < \frac{3(1+w)}{2(2+3w)} \ln\frac{H_0}{T_0} + \frac{1-3w}{2(2+3w)}\ln \frac{\mpl}{T_\rh} \,.
\ee
The right hand side of \eqref{Nk3} is monotonically decreasing function of $w$ for the range of our interest, 
so the inequality is easier to satisfy for a minimal $w$, i.e. the near-critical expansion after inflation with $w\approx-1/3$. This gives
\be\label{bound2}
\frac{H_k}{\mpl}<\frac{\mpl H_0}{T_\rh T_0}\approx64\frac{T_0}{T_\rh}.
\ee
which is consistent with the simple bound \eqref{bound1}.  

This bound can also be written in the tensor-to-scalar ratio $r$ defined as the ratio between the amplitude of the power spectrum for the primordial tensor perturbation $\mathcal{P}_T$ and that of the curvature perturbation $\mathcal{P_R}$, which is measured to a very high accuracy. We have
\be\label{Hk}
\left(\frac{H_k}{\mpl}\right)^2=\frac{\pi^2}{2}r\mathcal{P}_\mathcal{R}(k) \, .
\ee
We use the value of $\mathcal{P_R}$ on the pivot scale $k_p=0.02\text{Mpc}^{-1}$ and neglect the small running to the comoving scale we consider, $k_\text{DE}$, because $\mathcal{P}_T$ and $\mathcal{P}_\mathcal{R}$ are nearly scale-invariant on CMB scales. Therefore we have the bound on $r$:
\be\label{bound3}
r\lesssim2\times10^{-8}\left(\frac{1\text{MeV}}{T_\rh}\right)^2.
\ee
We normalize the reheating temperature to its lowest possible value $1$MeV from the requirement of big bang nucleosynthesis~\cite{Kawasaki:1999na,Kawasaki:2000en,Hannestad:2004px,Hasegawa:2019jsa}. The current and future CMB B-mode polarization experiments, like BICEP/Keck~\cite{Ade:2018gkx}, AliCPT~\cite{Li:2017drr}, and LiteBIRD~\cite{Matsumura:2016sri}, are possible to detect $r$ as low as $10^{-3}$, which is still far from \eqref{bound3}. However, our universal bound \eqref{bound3} is a great improvement of the BBLV bound $r<10^{-30}$, which can be reproduced from our universal bound by setting $T_\rh\sim\sqrt{\mpl H_k}$. And this tensor-to-scalr ratio is typical in some supergravity- or string-inspired inflation models, for instance brane inflation~\cite{Dvali:1998pa}, KKLT inflation~\cite{Kachru:2003aw}, K\"ahler moduli inflation~\cite{Conlon:2005jm}, punctuated inflation~\cite{Jain:2009pm}, etc. 

\begin{figure}[htbp]
\begin{center}
\includegraphics[trim={2cm 1cm 1cm 1cm}, clip, width=0.5\textwidth]{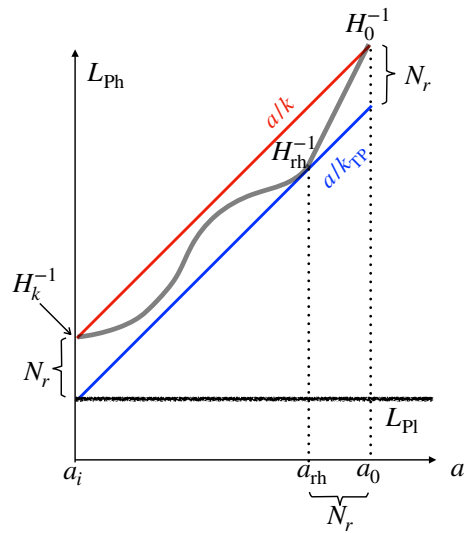}
\caption{``Smoothed'' multi-stage inflation which saturates the upper bound of \eqref{bound1}. It can be clearly seen that only the reheating temperature is relevant for predicting $H_k$, as long as there is no violation of \eqref{condition}.}
\label{fig:smooth}
\end{center}
\end{figure}

\section{Conclusion and Discussion}\label{sec:con}
In this paper, based on the recently proposed Trans-Planckian Censorship Conjecture (TCC), we found a universal upper bound on the inflationary Hubble expansion rate, which is inversely proportional to the reheating temperature: $H_k/\mpl\lesssim T_0/T_\rh$. In constructing a model that saturates the upper bound,  we consider an oscillatory stage after inflation with a near-critical equation-of-state parameter $w\approx-1/3$, and a low reheating temperature $T_\rh<10^8\text{GeV}$. For the lowest reheating temperature $1\text{MeV}$, the upper bound on the tensor-to-scalar ratio $r$ at the CMB scales is around $10^{-8}$, which can be realized in, e.g. brane inflation. 

This near-critical expansion to reach the upper limit seems \textit{ad hoc}. However,  based on the same idea, we can find a group of models that have the same prediction. First, we notice that there are two relevant scales: the scale of the horizon exit of the CMB scales which is fixed by the CMB observations, and the reheating scale which is constrained by the big bang nucleosynthesis. Between these two scales, we can have arbitrary intermediate stages of inflation or decelerated expansion, just keep in mind not to violate the TCC at any moment. This is equivalent to require the Hubble expansion rate to satisfy 
\be\label{condition}
a/k_\text{DE}>H_\text{inter}^{-1}>a/k_\text{rh}
\ee
between $a_i$ and $a_\rh$, where $k_\rh\sim a_0T_0T_\rh/\mpl$ is the comoving wavenumber which is tangent to the horizon at the reheating. See Fig.~\ref{fig:smooth}. 
This is the extension of the inflation with multiple stages, with a smoothly varying Hubble expansion rate. 
The condition \eqref{condition} for the intermediate stage is important, as when $a/k_\text{DE}<H^{-1}$, we may have some unwanted features on CMB anisotropies, similar to some previous works on the large scale anomalies like \cite{Polarski:1992dq,Polarski:1995zn,Langlois:1999dw}. When $H^{-1}<a/k_\text{rh}$, there will be another trans-Planckian mode with larger wavenumber, which makes the prediction of $H_k/\mpl$ even smaller than $T_0/T_\rh$. 

\begin{figure}[htbp]
\begin{center}
\includegraphics[trim={0cm 1cm 1cm 1cm}, clip, width=0.5\textwidth]{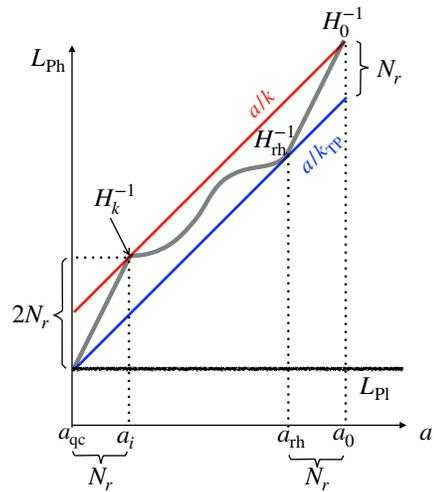}
\caption{When there is a pre-inflationary stage with equation-of-state parameter $w_\text{pi}$, the TCC induced bound on $H_k$ will change. In this figure we draw the $w_\text{pi}=1/3$ case, which corresponds to a pre-inflationary radiation dominated era. The upper limit of Hubble expansion rate in this figure can be easily derived by the trigonometry as  $H_k/\mpl=(T_0/T_\rh)^2$.}
\label{fig:pre}
\end{center}
\end{figure}

A key assumption of the BBLV bound and our bound \eqref{bound1} is that the universe starts at $a_i$, and the initial condition for the quantum fluctuations is set at that moment. However, if there is a pre-inflationary stage to set the initial condition of inflation, as is studied for instance in Refs.~\cite{Bhattacharya:2005wn,Powell:2006yg,Wang:2007ws,Das:2014ffa}, our upper bound \eqref{bound1} should be revised by adding a power of $T_0/T_\rh$. For this purpose, we suppose that the equation-of-state parameter during the pre-inflationary stage is $w_\text{pi}>-1/3$, and draw its spacetime diagram as in Fig.~\ref{fig:pre}. The trans-Planckian mode corresponds to the one tangent to the Hubble horizon at reheating, but now it originates from an early pre-inflationary stage. Simple trigonometry tells us that the bound \eqref{bound1} should be revised to 
\be\label{bound4}
\frac{H_k}{\mpl}\lesssim\left(\frac{T_0}{T_\rh}\right)^{1+\frac2{1+3w_\text{pi}}},
\ee
which becomes more stringent than our previous bound \eqref{bound1}, and even stronger than the BBLV bound \eqref{BBLV} for the lowest reheating temperature if $w_\text{pi}<1/3$. Therefore we concluded that \eqref{bound1} should be comprehended as a conservative bound, which can be saturated only for the case when the inflationary universe is created right at $a_i$. 

If the Hubble expansion rate does evolve as shown in Fig.~\ref{fig:smooth} and Fig.~\ref{fig:pre} and satisfy \eqref{condition}, we must have some modes with intermediate wavelengths which repeatedly go inside and outside the Hubble horizon. These modes may give us some additional features and/or enhancement of the power spectra of the curvature perturbations and tensor perturbations on specific scales. We already know that such a scenario can not enhance primordial gravitational waves~\cite{Pi:2019ihn}. However, enhancement of curvature perturbations may increase the formation of substantial primordial black holes, which may be the candidate for dark matter~\cite{GarciaBellido:1996qt,Carr:2009jm,Frampton:2010sw,Carr:2016drx,Green:2004wb}, or the black hole binaries detected by LIGO~\cite{Sasaki:2016jop,Clesse:2016vqa,Bird:2016dcv}. We will study this interesting topic in the future.

We now would like to comment more on the physical implications of the TCC. 
The TCC appears to be conceptually similar to the so-called the de Sitter entropy bound conjecture~\cite{ArkaniHamed:2007ky}, which states that $N_{\rm obs}\lesssim S_{\rm end}$. Here, $N_{\rm obs}$ is the observable number of $e$-foldings, $S_{\rm end} = \pi/(G H_{\rm end}^2)$ is the de Sitter entropy at the end of inflation and $H_{\rm end}$ is the Hubble expansion rate at the end of inflation. Since the bound involves the observable number of $e$-foldings but not the total number, the de Sitter entropy conjecture for inflationary models can be significantly weakened by the presence of dark energy in the late time universe~\cite{Jazayeri:2016jav}. In particular, it permits an infinitely long period of accelerating expansion in the late universe. This is because inflationary modes generated in the early universe never enters the horizon during the dark energy dominated era. On the other hand, the TCC forbids any modes that were previously trans-Planckian from becoming superhorizon, irrespectively of whether those modes are observable or not. For this reason, the TCC even restricts the number of $e$-foldings during the dark energy dominated era in the late universe. For example, for the model shown in Fig.~\ref{fig:pre} the TCC tells that the number of $e$-foldings for the late time acceleration cannot exceed $N_r$. Another difference between the two conjectures is in their implications. By definition, the de Sitter entropy bound is satisfied at early time of the universe but can be violated at late time. If the de Sitter entropy bound is violated at a critical time, which we may call the cosmological Page time, and if the conjecture is true then the prediction of semi-classical effective field theory should start to deviate from the correct result by $\mathcal{O}(1)$~\cite{Jazayeri:2016jav}. However, the violation at the cosmological Page time does not imply the breakdown of the effective theory before that time. On the other hand, in the case of the TCC, any effective theories that do not satisfy the condition stated in it at one time are considered to be in the swampland all the time. In this sense, the TCC is stronger than the de Sitter entropy bound conjecture. Indeed, the only model of inflation that the de Sitter entropy bound conjecture claimed to rule out is ghost inflation~\cite{ArkaniHamed:2003uz}, yet this claim was later weakened~\cite{Jazayeri:2016jav}. On the other hand, the TCC potentially rules out many models of inflation with $r\gtrsim10^{-8}$, if it is true. 

In the end, we would like to comment that the TCC is
related to but actually different from the traditional \textit{trans-Planckian problem}. The initial condition of the quantum fluctuations during inflation must be set deep inside the horizon. For the sub-Planckian fluctuations which are stretched out of horizon during inflation, there may be possible corrections from the quantum gravity, which is described by the trans-Planckian problem~\cite{Martin:2000xs,Brandenberger:2000wr,Niemeyer:2000eh,Starobinsky:2001kn,Easther:2001fi,Danielsson:2002kx,Easther:2002xe}. In inflation models compatible with TCC, however, as the initial conditions are set at the beginning of inflation, and the sub-Planckian wavelengths at that moment will stay subhorizon forever, the trans-Planckian problem is absent.

\subsection*{Acknowledgements}
We thank Rong-Gen Cai, Shao-Jiang Wang, and Yi Zhang for useful discussions. 
The work of S. Mukohyama was supported by Japan Society for the Promotion of Science (JSPS) Grants-in-Aid for Scientific Research (KAKENHI) No. 17H02890, No. 17H06359. SP was supported by the MEXT/JSPS KAKENHI No. 15H05888, and JSPS Grant-in-Aid for Early-Career Scientists No.20K14461. S. Mukohyama and SP were also partially supported by the World Premier International Research Center Initiative (WPI Initiative), MEXT, Japan. YLZ was supported by Grant-in-Aid for JSPS international research fellow(18F18315).

\end{document}